# Improvement on 'Secure multi-party quantum summation based on quantum Fourier transform'

Jun Gu[1], Tzonelih Hwang[*]

*Department of Computer Science and Information Engineering, National Cheng Kung University, No. 1, University Rd., Tainan City, 70101, Taiwan, R.O.C.*

[1] isgujun@163.com

[*] hwangtl@csie.ncku.edu.tw

[*]**Corresponding Author:**

Tzonelih Hwang

Distinguished Professor

Department of Computer Science and Information Engineering,

National Cheng Kung University,

No. 1, University Rd.,

Tainan City, 70101, Taiwan, R.O.C.

Email: hwangtl@csie.ncku.edu.tw

TEL: +886-6-2757575 ext. 62524

# Abstract

Recently, Yang et al. (Quantum Inf Process:17:129, 2018) proposed a secure multi-party quantum summation protocol allowing the involved participants to sum their secrets privately. They claimed that the proposed protocol can prevent each participant's secret from being known by others. However, this study shows that the participant who prepares the initial quantum states can obtain other participants' secrets with an inverse quantum Fourier transform attack. A modification is then proposed here to solve this problem.

**Keywords** Quantum summation. Inverse quantum Fourier transform attack. Secure quantum computation.

## 1. Introduction

The task of secure multi-party quantum summation (SMQS) is helping the involved participants to obtain the summation of their secrets, and at the same time, the participants' secrets can be kept in privacy. In 2007, Du et al. [1] designed an SMQS protocol based on single photons. Then, to improve the efficiency of Du et al.'s SMQS protocol, Chen et al. [2] designed an efficient SMQS protocol with GHZ states, Zhang et al. [3] proposed a high-capacity SMQS protocol where both of the polarization and the spatial-mode degrees of freedom are used. In addition, several other SMQS protocols have been proposed [4-8].

Recently, Yang et al. [9] proposed an SMQS protocol based on quantum Fourier transform and claimed that their protocol can avoid both of the outside attacks and the inside attacks. However, this study shows that the participant who generates the initial particles can use an inverse quantum Fourier transform attack to obtain other participants' secrets. Then, a simple solution is hence proposed here.

The rest of this paper is organized as follows. Section 2 briefly reviews Yang et al.'s SMQS protocol. Section 3 first shows the details of the inverse quantum Fourier transform attack on Yang et al.'s protocol and then proposes a modified method. At last, a brief conclusion is given in Section 4.

## 2. Brief review of Yang et al.'s SMQS protocol [9]

Before reviewing Yang et al.'s SMQS protocol, it is necessary to briefly introduce some background first.

### 2.1 Background

In Yang et al.'s protocol, the $d$-level transformation operation $U_k$, $d$-level quantum Fourier transform operation $QFT$ and their applications on the $d$-level $n$-particle entangled state $|\omega\rangle_{a_1 a_2 \cdots a_n}$ are used. Assume $|r\rangle (r \in \{0,1,\cdots,d-1\})$ is a $d$-level single particle, then the $U_k$, $QFT$ and $|\omega\rangle_{a_1 a_2 \cdots a_n}$ can be respectively described as follows:

$$U_k |r\rangle = |k \oplus r\rangle \left(k \in \{0,1,\cdots,d-1\}\right) \tag{1}$$

$$QFT |r\rangle = \frac{1}{\sqrt{d}} \sum_{l=0}^{d-1} e^{\frac{2\pi i l r}{d}} |l\rangle \tag{2}$$

$$|\omega\rangle_{a_1 a_2 \cdots a_n} = \frac{1}{\sqrt{d}} \sum_{r=0}^{d-1} |r\rangle_{a_1} |r\rangle_{a_2} \cdots |r\rangle_{a_n} \tag{3}$$

Hence, if performing the operations $(U_{k_1} QFT) \otimes (U_{k_2} QFT) \otimes \cdots \otimes (U_{k_n} QFT)$ on the state $|\omega\rangle_{a_1 a_2 \cdots a_n}$, we can obtain the following formula:

$$(U_{k_1} QFT) \otimes (U_{k_2} QFT) \otimes \cdots \otimes (U_{k_n} QFT) |\omega\rangle_{a_1 a_2 \cdots a_n}$$

$$= d^{-\frac{n-1}{2}} \sum_{l_1+l_2+\cdots+l_n \equiv 0 (\bmod d)} |l_1 \oplus k_1\rangle_{a_1} |l_2 \oplus k_2\rangle_{a_2} \cdots |l_n \oplus k_n\rangle_{a_n} \tag{4}$$

where $(l_1 \oplus k_1) \oplus (l_2 \oplus k_2) \oplus \cdots \oplus (l_n \oplus k_n) = (k_1 + k_2 + \cdots + k_n) \bmod d$.

### 2.2 Yang et al.'s SMQS protocol

Suppose that there are $n$ participants $(P_1, P_2, \cdots, P_n)$ who want to obtain the summation of their private integer strings. Let $P_i$'s private integer string is $K_i = (k_i^1, k_i^2, \cdots, k_i^m)$ where $i \in \{1, 2, \cdots, n\}$ and $m$ is the length of $K_i$. Then Yang et al.'s SMQS protocol can be presented step by step as follows:

**Step 1** $P_1$ generates $m$ entangled states in $|\omega\rangle_{a_1 a_2 \cdots a_n}$ and picks out all the $i$th particles $a_i$ in $|\omega\rangle_{a_1 a_2 \cdots a_n}$ to construct $n$ ordered particle sequences $(S_1, S_2, \cdots, S_n)$ where $S_i = (a_i^1, a_i^2, \cdots, a_i^m)$. Subsequently, $P_1$ inserts enough decoy particles into $S_i (i \in \{2, 3, \cdots, n\})$ to obtain new sequences $S_i'$ where each decoy particle is randomly generated with $V_1$-basis or $V_2$-basis, here, $V_1 = \{|r\rangle (r \in \{0, 1, \cdots, d-1\})\}$ and $V_2 = \{QFT|r\rangle (r \in \{0, 1, \cdots, d-1\})\}$. Then $P_1$ sends $S_i'$ to $P_i$ and keeps $S_1$ in his/her hand.

**Step 2** Upon receiving $S_i'$, $P_i (i \in \{2, 3, \cdots, n\})$ and $P_1$ use the decoy particles to check whether there is an eavesdropper in the particles transmission process. If the error rate exceeds a predetermined value, they abort this protocol. Otherwise, $P_i (i \in \{2, 3, \cdots, n\})$ discards the decoy particles in $S_i'$ to get $S_i = (a_i^1, a_i^2, \cdots, a_i^m)$ and continues the next step.

**Step 3** Each participant $P_i (i \in \{1, 2, \cdots, n\})$ encodes his/her private integer $K_i$ on $S_i$ by performing $U_{k_i^j} QFT$ on the particle $a_i^j (j \in \{1, 2, \cdots, m\})$ and uses $V_1$-basis to measure all the particles to obtain the corresponding measurement results $R_i = (r_i^1, r_i^2, \cdots, r_i^m)$ where $r_i^j = k_i^j \oplus l_i^j$. Finally, $P_i (i \in \{2, 3, \cdots, n\})$ announces $R_i$ to $P_1$.

**Step 4** Upon obtaining all the $R_i$, according to formula (4), $P_1$ can get the summation of all the participants' private integer strings $K_i$ with $Sum = R_1 \oplus R_2 \oplus \cdots \oplus R_n$. In detail, $Sum = R_1 \oplus R_2 \oplus \cdots \oplus R_n = K_1 \oplus L_1 \oplus K_2$ $\oplus L_2 \oplus \cdots \oplus K_n \oplus L_n = K_1 \oplus K_2 \oplus \cdots \oplus K_n$. Then, $P_1$ announces *Sum* to $P_i \left( i \in \{2, 3, \cdots, n\} \right)$.

## 3. Attack and modification on Yang et al.'s SMQS protocol

Yang et al. [9] claimed that the above protocol can prevent each participant's private integer string $K_i \left( i \in \{1, 2, \cdots, n\} \right)$ from being known by others. However, this section shows that, in Yang et al.'s secure multi-party quantum summation protocol, $P_1$ can use an inverse quantum Fourier transform attack to obtain other participants' private integer strings $K_i \left( i \in \{2, 3, \cdots, n\} \right)$. Then, to solve this problem, a modification is proposed.

### 3.1 The Inverse Quantum Fourier Transform Attack

In Step 2, the participants just use decoy particles to check whether there is an eavesdropper in the particle transmission processes, but do not check the correctness of the initial states $|\omega\rangle_{a_1 a_2 \cdots a_n}$. Hence, if $P_1$ is a malicious participant and he/she sends fake initial states to other participants in Step 1, he/she can obtain other participants' private integer strings. That is, in Step 1, $P_1$ generates the fake single particles $QFT^{-1}|r\rangle$ instead of the states $|\omega\rangle_{a_1 a_2 \cdots a_n}$ where $QFT^{-1}$ is the $d$-level inverse quantum Fourier transform operation and $QFT^{-1}|r\rangle = \frac{1}{\sqrt{d}} \sum_{x=0}^{d-1} e^{\frac{-2\pi i x r}{d}} |x\rangle$. Then, he/she sends these particles with decoy particles to $P_i$. After the eavesdropper detection with decoy particles in Step 2, $P_i$ will encode his/her private integer strings by performing $U_{k_i^j} QFT$ on $QFT^{-1}|r\rangle$ in Step 3. The result of $U_{k_i^j} QFT QFT^{-1}|r\rangle$ is $|r \oplus k_i^j\rangle$ and the details are shown as follows:

$$\begin{aligned}
&U_{k_i^j}QFTQFT^{-1}|r\rangle \\
&=U_{k_i^j}QFT\frac{1}{\sqrt{d}}\sum_{x=0}^{d-1}e^{\frac{-2\pi irx}{d}}|x\rangle \\
&=U_{k_i^j}\frac{1}{\sqrt{d}}\sum_{l=0}^{d-1}e^{\frac{2\pi ixl}{d}}\frac{1}{\sqrt{d}}\sum_{x=0}^{d-1}e^{\frac{-2\pi irx}{d}}|l\rangle \\
&=U_{k_i^j}\frac{1}{d}\sum_{x=0}^{d-1}\sum_{l=0}^{d-1}e^{\frac{2\pi i(l-r)x}{d}}|l\rangle \\
&=U_{k_i^j}\left(\frac{1}{d}\sum_{x=0}^{d-1}e^{\frac{2\pi i(r-r)x}{d}}|r\rangle+\frac{1}{d}\sum_{l=0\wedge l\neq r}^{d-1}\sum_{x=0}^{d-1}e^{\frac{2\pi i(l-r)x}{d}}|l\rangle\right) \\
&=U_{k_i^j}\left(\frac{1}{d}\sum_{x=0}^{d-1}|r\rangle+\frac{1}{d}\sum_{l=0\wedge l\neq r}^{d-1}0|l\rangle\right) \\
&=U_{k_i^j}|r\rangle \\
&=|r\oplus k_i^j\rangle
\end{aligned} \qquad (5)$$

Subsequently, $P_i$ uses $V_1$-basis to measure all the particles and announces the measurement results $R_i=(r\oplus k_i^1, r\oplus k_i^2,\cdots,r\oplus k_i^m)$. In Step 4, after $P_1$ obtains all the $R_i$, he/she can use the formula $r\oplus k_i^j\oplus r=k_i^j$ to get all the private integer strings $K_i=(k_i^1,k_i^2,\cdots,k_i^m)$.

For example, assume that there are three participants $(P_1,P_2,P_3)$ who want to sum their private integer strings $(K_1,K_2,K_3)$. For simplify, there is just one integer in each string, $K_1=4, K_2=5, K_3=6$. In Step 1, if $P_1$ sends the fake state $QFT^{-1}|2\rangle$ to $P_2$ and $P_3$. Then, after $P_2$ performs $U_5QFT$ on $QFT^{-1}|2\rangle$, the fake state will be transformed to $|7\rangle$ and the measurement result $R_2$ will be $7$. Similarly, $P_3$ will obtain the measurement result $R_3=8$. Upon receiving $R_2=7$ and $R_3=8$, $P_1$ will know that $P_2$'s private integer is $7-2=5$ and $P_3$'s private integer is $8-2=6$. Hence, $P_1$ can obtain other participants' private integer by the inverse quantum Fourier transform attack.

## 3.2 Modification on Yang et al.'s SMQS protocol

Because the participants do not check the correctness of the initial states in the protocol, $P_1$ can use the inverse quantum Fourier transform attack to get other participants' private integer strings. Thus, if a process for checking whether the initial states $|\omega\rangle_{a_1 a_2 \cdots a_n}$ are correct or not is added, then the problem can be solved. The checking method of $|\omega\rangle_{a_1 a_2 \cdots a_n}$ used is based on [10]. The modified version is as follows.

**Step 1*** $P_1$ generates $m+\eta$ entangled states in $|\omega\rangle_{a_1 a_2 \cdots a_n}$ and constructs $n$ ordered particle sequences $S_i = (a_i^1, a_i^2, \cdots, a_i^{m+\eta})$ similarly. The rest parts of Step 1* are the same as the Step 1 in Section 2.

**Step 2*** (the same as **Step 2** in Section 2.)

**Step 3*** For checking the correctness of the initial states $|\omega\rangle_{a_1 a_2 \cdots a_n}$, each participant $P_i (i \in \{2, 3, \cdots, n\})$ randomly picks out $\frac{\eta}{n-1}$ states from all the initial states as checking states. For each picked state, $P_i$ randomly chooses a measurement basis from $\{V_1, V_2\}$ and announces the checking state position and the corresponding chosen basis. Then, each participant performs a $QFT$ operation on the corresponding particle and uses the chosen basis to measure it. After that, $P_1$ first announces his measurement result, and then all the other participants announce their measurement results. According to Eq. (4) where $k_i = 0$, if the chosen basis is $V_1$, the summation of all the measurement results will be 0 when it is divided by $d$. And if the chosen basis is $V_2$, according to

$$QFT^{\otimes n} |\omega\rangle_{a_1 a_2 \cdots a_n} = \frac{1}{\sqrt{d}} \sum_{r=0}^{d-1} QFT|r\rangle_{a_1} QFT|r\rangle_{a_2} \cdots QFT|r\rangle_{a_n} \quad ,\quad \text{all the}$$

measurement results will be the same. Hence, if all the corresponding measurement results are correct, they discard these checking states and continue the next step. Otherwise, they abort this protocol.

**Step 4*** and **Step 5*** are the same as **Step 3** and **Step 4** in Section 2, respectively.

With this checking process, the participants can ensure that the shared initial states are $|\omega\rangle_{a_1 a_2 \cdots a_n}$ [10]. Hence, the inverse quantum Fourier transform attack mentioned earlier can be avoided.

## 4. Conclusions

Yang et al. proposed an SMQS protocol based on quantum Fourier transform operation. However, this study points out a loophole in the protocol. With the loophole, a malicious participant $P_1$ may obtain other participants' private integer strings with an inverse quantum Fourier transform attack. An improvement is hence proposed to avoid this loophole.

## Acknowledgement

We would like to thank the Ministry of Science and Technology of the Republic of China, Taiwan for partially supporting this research in finance under the Contract No. 107-2221-E-006 -077 -.